\documentstyle[prb,aps,epsf]{revtex}

\begin{document}
\draft

\advance\textheight by 0.5in \advance\topmargin by -0.in

\twocolumn[\hsize\textwidth\columnwidth\hsize\csname@twocolumnfalse\endcsname{ }

\title{Persistent current and correlation effects in carbon nanotubes}
\author{Arkadi A. Odintsov$^{1,2}$, Wolter Smit$^{1}$, and Hideo Yoshioka$^{1,3}$}
\address{
$^{1}$Department of Applied Physics, Delft University of Technology,
2628 CJ Delft, The Netherlands \\
$^{2}$Nuclear Physics Institute, Moscow
State University, Moscow 119899 GSP, Russia\\
$^{3}$Department of Physics, Nagoya University,
Nagoya 464-01, Japan. \\
}
\date{\today}
\maketitle

\begin{abstract}
The persistent current of interacting electrons in toroidal single-wall
carbon nanotubes is evaluated within Haldane's concept of topological
excitations. The overall pattern of the persistent current corresponds to
the constant interaction model, whereas the fine structure stems from the
electronic exchange correlations. 
\end{abstract}
\vskip -0.5 truein
]


The recent breakthrough in the synthesis of a new generation of quantum
wires - single-wall carbon nanotubes (SWNTs) \cite{Thess} and the subsequent
observation \cite{Tans,Bockrath} of coherent electron transport in this
system have initiated a surge of experimental and theoretical activity (see
e.g. Refs. \cite{Tans2,Cobden,Krotov,Egger,Kane,YoshOdi}) . The
investigation of non-Fermi liquid correlation effects due to one-dimensional
nature of interacting electrons in SWNTs presents one of the main
challenges. The signatures of such correlations are often masked by the
charging effects. Nevertheless, very recent experimental results \cite{Tans2}
on the spin structure of the ground state suggest the interpretation in
terms of electron correlations.

Generically, carbon nanotubes have linear or curved shape. Recently Liu et.
al. \cite{Liu} have observed individual circular SWNTs and ropes of such
nanotubes. The experimental data suggests uniform widths of SWNTs and does
not display the presence of their ends. For this reason it is plausible that
circular SWNTs have the topology of a torus. One can test this conjecture by
measuring the response of circular SWNTs to a weak perpendicular magnetic
field. Since SWNTs are typically almost free from defects \cite{Thess}, the
presence of delocalized electronic states in the system should result in a
persistent current. To provide theoretical support for future experiments,
we analyze the persistent current in toroidal SWNTs (TNTs).

We are aware of two recent papers \cite{Haddon,Lin} where the persistent
current in TNTs has been computed from first principles within the
single-particle scheme. The results are either related to the specific
fullerenes ($C_{576}$ carbon toroids \cite{Haddon}), or limited to the
special case of half-filling \cite{Lin}. Moreover, both works ignore
electron correlations due to the Coulomb interaction, whose observable
signatures in the pattern of the persistent current are investigated in this
Letter.

We employ bosonization formalism \cite{Gogolin} which has proven to be
effective for analysis of persistent currents in various one-dimensional
models \cite{pctheory}. First, we extend the bosonization scheme \cite
{Egger,YoshOdi,comm2} (see also \cite{Kane}) to the case of TNTs of the
''armchair'' $(N,N)$ type. The Fermi operators $\Psi _{s}$ for electrons
with spin $s=\pm $ can be expanded near the two crossing (or Fermi) points $%
\alpha K$ ($\alpha =\pm $, $K=4\pi /3a$) of the energy spectrum into right- (%
$d=+$) and left-moving ($d=-$) components, $\Psi _{s}(x)=\sum_{\alpha
d}e^{i\alpha Kx}\psi _{asd}(x).$ The periodicity of the electronic fields $%
\Psi _{s}(x+L)=\Psi _{s}(x)$ results in the following boundary condition for
the slowly-varying parts, $\psi _{asd}(x+L)=\psi _{asd}(x)e^{2\pi i\alpha
p/3}$ where $p=0,\pm 1$ parametrizes the number $L/a$ of elementary cells
along the nanotube of length $L$, $L/a=3n+p$ ($n$ is an integer).

Bosonization allows one to express the Fermi operators $\psi
_{asd}(x)\propto e^{i(\phi _{\alpha s}+d\theta _{\alpha s})}$ in terms of
the bosonic fields $\theta _{\alpha s}$ and $\phi _{\alpha s}$ obeying the
commutation rules $\left[ \theta _{\alpha s}(x),\phi _{\alpha ^{\prime
}s^{\prime }}(x^{\prime })\right] =(\pi i/2)\mbox{sign}(x-x^{\prime })\delta
_{\alpha \alpha ^{\prime }}\delta _{ss^{\prime }}$. The fields can be
decomposed into topological parts and non-zero bosonic modes $\tilde{\theta}%
_{\alpha s}$, $\tilde{\phi}_{\alpha s}$: 
\begin{eqnarray}
\theta _{\alpha s}(x) &=&\theta _{\alpha s}^{(0)}+(N_{\alpha s}+1)\pi x/L+%
\tilde{\theta}_{\alpha s}(x),  \label{theta} \\
\phi _{\alpha s}(x) &=&\phi _{\alpha s}^{(0)}+(J_{\alpha s}+2p\alpha /3)\pi
x/L+\tilde{\phi}_{\alpha s}(x).  \label{phi}
\end{eqnarray}
The pairs of the action-angle operators $J_{\alpha s},\theta _{\alpha
s}^{(0)}$ and $N_{\alpha s},\phi _{\alpha s}^{(0)}$ satisfy the canonical
commutation relation, $\left[ N_{\alpha s},\phi _{\alpha ^{\prime }s^{\prime
}}^{(0)}\right] =\left[ J_{\alpha s},\theta _{\alpha ^{\prime }s^{\prime
}}^{(0)}\right] =-i\delta _{\alpha \alpha ^{\prime }}\delta _{ss^{\prime }}.$
The topological excitations $N_{\alpha s}$, $J_{\alpha s}$ are simply
related to the numbers $M_{\alpha sd}=(N_{\alpha s}+dJ_{\alpha s})/2$ of
excess electrons at the branch $\alpha $, $s$, $d$ of the energy spectrum,
see inset of Fig.~1a. Since $M_{\alpha sd}$ are integers, the sum $N_{\alpha
s}+J_{\alpha s}$ must be even (formally this topological constraint follows
from the boundary condition on $\psi $-operators).

\smallskip Following Refs. \cite{Egger,Kane,YoshOdi} we introduce bosonic
fields $\theta _{\delta \nu }(x)$ and $\phi _{\delta \nu }(x)$ describing
the charge $\nu =+$ and spin $\nu =-$ excitations in the symmetric $\delta
=+ $ and antisymmetric $\delta =-$ modes, 
\begin{equation}
O_{\delta \nu }=[O_{++}+\nu O_{+-}+\delta (O_{-+}+\nu O_{--})]/2,
\label{transform}
\end{equation}
where $O=\theta $ ($\tilde{\theta},\theta ^{(0)},N$) or $\phi $ ($\tilde{\phi%
},\phi ^{(0)},J$) and the indices in the r.h.s. correspond to $\alpha $, $s$%
. The new fields,

\begin{eqnarray}
\theta _{\delta \nu }(x) &=&\theta _{\delta \nu }^{(0)}+(N_{\delta \nu
}+2\delta _{\delta +}\delta _{\nu +})\pi x/L+\tilde{\theta}_{\delta \nu }(x),
\label{thetanew} \\
\phi _{\delta \nu }(x) &=&\phi _{\delta \nu }^{(0)}+(J_{\delta \nu
}+(4/3)p\delta _{\delta -}\delta _{\nu +})\pi x/L+\tilde{\phi}_{\delta \nu
}(x),  \label{phinew}
\end{eqnarray}
satisfy the commutation relations, $\left[ \theta _{\delta \nu }(x),\phi
_{\delta ^{\prime }\nu ^{\prime }}(x^{\prime })\right] =(\pi i/2)\mbox{sign}%
(x-x^{\prime })\delta _{\delta \delta ^{\prime }}\delta _{\nu \nu ^{\prime
}} $, $\left[ N_{\delta \nu },\phi _{\delta ^{\prime }\nu ^{\prime
}}^{(0)}\right] =\left[ J_{\delta \nu },\theta _{\delta ^{\prime }\nu
^{\prime }}^{(0)}\right] =-i\delta _{\delta \delta ^{\prime }}\delta _{\nu
\nu ^{\prime }}$. The topological constraint for $N_{\alpha s}+J_{\alpha s}$
implies that $\sum_{\delta \nu }N_{\delta \nu }$, $\sum_{\delta \nu
}J_{\delta \nu }$, $\sum_{\nu }N_{\delta \nu }+J_{\delta \nu }$, $%
\sum_{\delta }N_{\delta \nu }+J_{\delta \nu }$, all must be even, whereas $%
\sum_{\delta \nu }N_{\delta \nu }+J_{\delta \nu }=0\mbox{ mod }4$. In
addition, the new topological numbers should be either all integer or all
half-integer. Note that $N_{tot}=2N_{++}+4$, $2N_{++}=\sum_{\alpha
sd}M_{\alpha sd}$ is the total number of extra electrons in the system \cite
{comm3}, whereas $2J_{++}=\sum_{\alpha sd}dM_{\alpha sd}$ is the difference
in numbers of right- and left-movers.

We concentrate first on the Luttinger model-like term \cite
{Egger,YoshOdi,Kane} $H_{L}$ of the low-energy Hamiltonian $H=H_{L}+V$ of
TNTs, 
\begin{eqnarray}
H_{L}=\sum_{\delta \nu } &&\int_{-L/2}^{L/2}\frac{dx}{2\pi }\left\{ \frac{%
v_{\delta \nu }}{K_{\delta \nu }}\left( \nabla \theta _{\delta \nu }-\frac{%
2K_{\delta \nu }}{v_{\delta \nu }}\mu \delta _{\delta +}\delta _{\nu
+}\right) ^{2}\right.  \nonumber \\
&&\left. +v_{\delta \nu }K_{\delta \nu }\left( \nabla \phi _{\delta \nu }-%
\frac{2e}{c}A\delta _{\delta +}\delta _{\nu +}\right) ^{2}\right\} ,
\label{H}
\end{eqnarray}
where the standard interaction parameters $K_{\delta \nu }$ and velocities $%
v_{\delta \nu }$ of excitations are introduced, so that $K_{\delta \nu }=1$
for non-interacting electrons. The electro-chemical potential $\mu $ of an
electronic reservoir controls the charge density $\rho =2e\nabla \theta
_{++}/\pi $. The vector potential $A$ of an external magnetic field is
coupled to the persistent current $I=-2e\dot{\theta}_{++}/\pi $. We will
neglect the Zeeman term $\mu _{B}H$ which is a factor $\sim a/L$ smaller
than the energy scale $v_{F}/L$ of interest.

The Hamiltonian $H_{L}$ splits into the bosonic part $H_{b}$ (which has the
form (\ref{H}) with $\theta \to \tilde{\theta}$, $\phi \to \tilde{\phi}$,
and $\mu =A=0$) and the topological part, 
\begin{eqnarray}
H_{t} &=&\frac{\pi }{2L}\sum_{\delta \nu }\frac{v_{\delta \nu }}{K_{\delta
\nu }}\left[ N_{\delta \nu }-4\left( f_{\mu }-\frac{1}{2}\right) \delta
_{\delta +}\delta _{\nu +}\right] ^{2}  \nonumber \\
&+&v_{\delta \nu }K_{\delta \nu }\left[ J_{\delta \nu }-4f_{\Phi }\delta
_{\delta +}\delta _{\nu +}+\frac{4p}{3}\delta _{\delta -}\delta _{\nu
+}\right] ^{2}.  \label{Htop}
\end{eqnarray}
Here $f_{\Phi }=\Phi /\Phi _{0}$ is the magnetic flux $\Phi $ through the
TNT in units of the flux quanta $\Phi _{0}=2\pi \hbar c/e$, and $f_{\mu
}=(K_{++}L/2\pi v_{++})\mu $ is a normalized electro-chemical potential,
whose reference point corresponds to the crossing of the energy spectrum.
The increase of $f_{\mu }$ by one corresponds to the addition of an electron
to each branch $\alpha $, $s$, $d$ of the spectrum, so that the properties
of the system are periodic in $f_{\mu }$ with a period of one (the same
periodicity occurs in $f_{\Phi }$). The Hamiltonian (\ref{Htop}) shows
additional symmetries with respect to changes in sign of the
electro-chemical potential ($f_{\mu }\to -f_{\mu }$, $M_{\alpha sd}\to
-M_{\alpha s-d}-1$) or the magnetic flux ($f_{\Phi }\to -f_{\Phi }$, $%
M_{\alpha sd}\to M_{-\alpha s-d}$).

The persistent current $I=dF/d\Phi $ can be calculated by differentiating
the free energy ${F}$ of the system with respect to the magnetic flux $\Phi $%
, 
\begin{equation}
I=(ev_{++}K_{++}/L)(8f_{\Phi }-2\langle J_{++}\rangle ).  \label{Ip}
\end{equation}
Due to the symmetries of the Hamiltonian, the persistent current is an even
(odd) periodic function of $f_{\mu }$ ($f_{\Phi }$). At zero temperature the
average $\langle J_{++}\rangle $ is determined by the ground state, whose
map is given in Figs. 1,2. The persistent current (\ref{Ip}) shows saw-tooth
dependence on the magnetic flux and changes in a stepwise manner as a
function of the electro-chemical potential. The amplitude of the persistent
current is given by $I_{\max }=4ev_{++}K_{++}/L\approx 0.5$ $\mu $A, for $%
v_{++}K_{++}\simeq v_{F}\approx 8\times 10^{5}$ m/s and $L=1$ $\mu $m. This
value is by two orders of magnitude larger than the persistent current
measured in GaAs mesoscopic rings \cite{Mailly}.

\strut The unscreened long-range Coulomb interaction strongly influences the
forward scattering of electrons leading to a large stiffness of the
symmetric charge mode \cite{Egger,Kane}, $K_{++}\approx 0.2$. We will first
ignore the sublattice-dependent part of the forward scattering as well as
the backscattering of electrons so that \cite{Kane} $v_{++}=v_{F}/K_{++}$
and $K_{\delta \nu }=1$, $v_{\delta \nu }=v_{F}$ for the modes $(\delta \nu
)=(+-),(-+),(--)$. Within this approximation, the energy spectrum of the
topological Hamiltonian (\ref{Htop}) is given by the sum of the Coulomb and
single-particle energies, which corresponds to the constant interaction
model (see e.g. Ref. \cite{Cobden}).

The ground state configurations for TNTs with $p=0,1$ are shown in Fig. 1.
Due to the spin degeneracy, the eigenstates of the Hamiltonian (\ref{Htop})
are characterized by four topological numbers, $M_{\alpha
d}=\sum_{s}M_{\alpha sd}$, Fig. 1b. For the nanotubes with $p=0$ the states
of electrons moving in the same direction at $\alpha =\pm $ are degenerate.
The system can be described by the two numbers $M_{d}=\sum_{\alpha
}M_{\alpha d}$ of extra right- and left-movers, Fig. 1a.

Since the Coulomb interaction in SWNTs is strong ($K_{++}\ll 1$), the
electron number $N_{tot}$ is determined primarily by the electro-chemical
potential, although in narrow regions it can be controlled by magnetic flux
(Fig. 1). The slope of the ground state borders enables one to deduce the
value of the interaction constant $K_{++}$ from experimental data \cite
{Eggert}. \smallskip Generically, the changes of the ground state with the
magnetic flux correspond to the jumps of electrons between different
branches of the spectrum and occur at universal values of magnetic flux - $%
f_{\Phi }=0,1/2$ for $p=0$ and $f_{\Phi }=0,1/6,1/3,1/2$ for $p=\pm 1$. In
particular, the jump of an electron at zero flux causes the paramagnetic
response of TNT. Such paramagnetic ground states occur if $N_{tot}\neq 4$ $%
\mathop{\rm mod}%
$ $8,$ for $p=0$, and $N_{tot}\neq 0$ $%
\mathop{\rm mod}%
$ $4$, for $p=\pm 1$ (see Fig. 1). Otherwise, the ground state is
diamagnetic.

The sublattice-dependent part of the forward scattering and backscattering
of electrons lead to the appearance of an essentially non-Luttinger term $%
V=V_{f}+V_{b}$ in the Hamiltonian and the renormalization of the parameters $%
K_{\delta \nu }$, $v_{\delta \nu }$ in Eq. (\ref{H}). The Luttinger and
non-Luttinger parts of the Hamiltonian describe intra- and interbranch
scattering of electrons respectively. The derivation of these terms from a
microscopic model has been discussed in Refs. \cite{YoshOdi,Egger,comm2} and
here we present only the results for a generic case away from half-filling.
The non-Luttinger terms are given by

\begin{eqnarray}
V_{f} &=&\frac{\Delta V(0)}{2\pi ^{2}\tilde{a}^{2}}\int_{-L/2}^{L/2}dx\{\cos
2\theta _{+-}\cos 2\theta _{--}  \label{Vf} \\
&&-\cos 2\theta _{-+}\cos 2\theta _{--}-\cos 2\theta _{-+}\cos 2\theta
_{+-}\},  \nonumber \\
V_{b} &=&\frac{1}{\pi ^{2}\tilde{a}^{2}}\int_{-L/2}^{L/2}dx\{\bar{V}(2K)\cos
2\theta _{+-}\cos 2\phi _{--}  \label{Vb} \\
&&+\frac{\Delta V(2K)}{2}[\cos 2\theta _{-+}\cos 2\theta _{--}+\cos 2\theta
_{-+}\cos 2\phi _{--}  \nonumber \\
&&\hspace{1.75cm}+\cos 2\theta _{--}\cos 2\phi _{--}]\},  \nonumber
\end{eqnarray}
where the matrix elements $\bar{V}(q)=[V_{+}(q)+V_{-}(q)]/2$, $\Delta
V(q)=V_{+}(q)-V_{-}(q)$ are related to the electron scattering amplitudes $%
V_{p}(q)$ given below, and $\tilde{a}\sim 1/K$ is the standard ultraviolet
cutoff in the Luttinger model. The interaction constants $K_{\delta \nu }=%
\sqrt{B_{\delta \nu }/A_{\delta \nu }}$, and velocities of excitations $%
v_{\delta \nu }=v_{F}\sqrt{A_{\delta \nu }B_{\delta \nu }}$ (\ref{H}) are
defined by \cite{YoshOdi} 
\begin{figure}[hbt]
\setlength{\unitlength}{1.0cm}
  \begin{picture}(8.5,12.5)
    \put(0.1,6.8){\epsfxsize=7.1cm\epsfbox{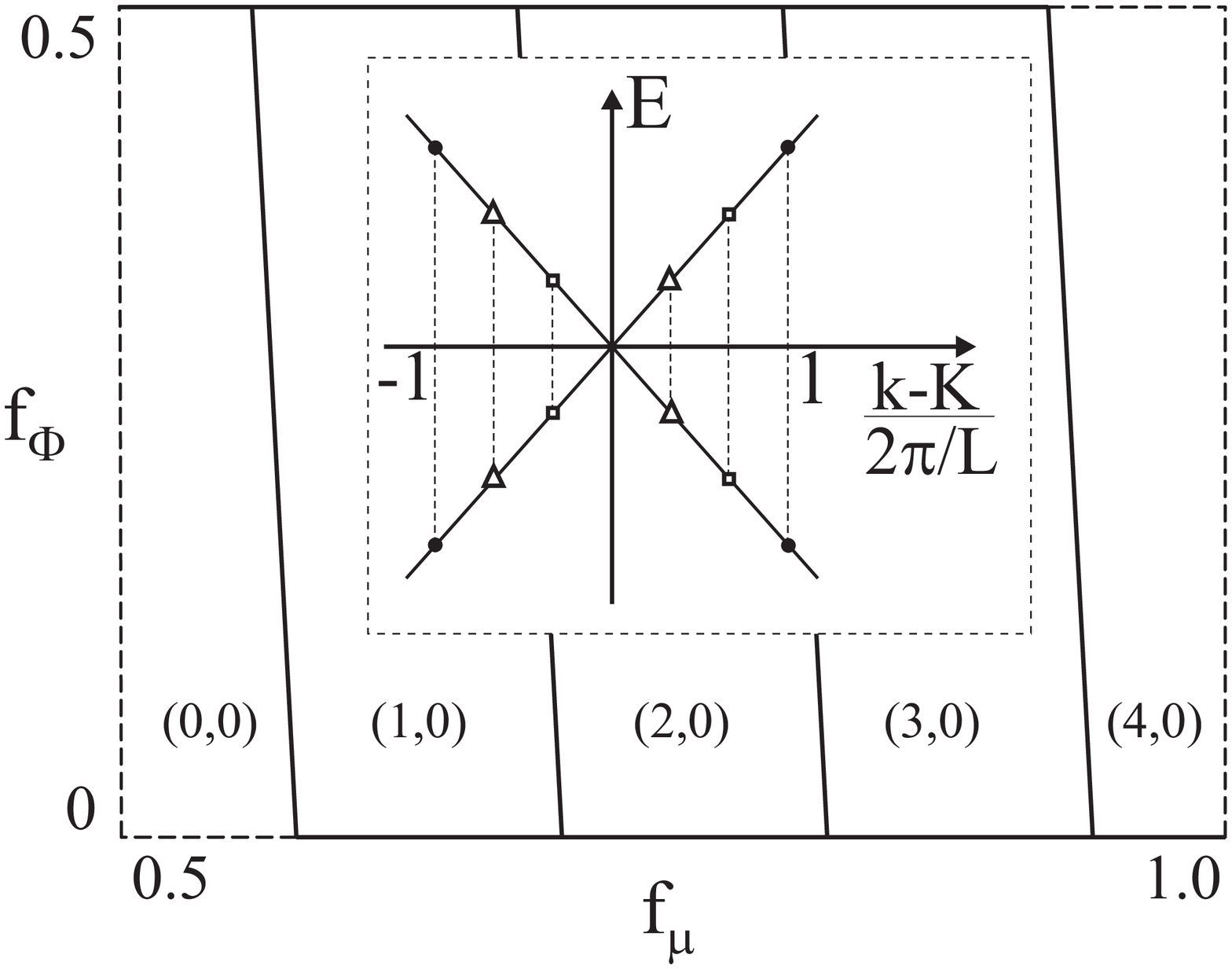}}
    \put(0.,1.0){\epsfxsize=8.0cm\epsfbox{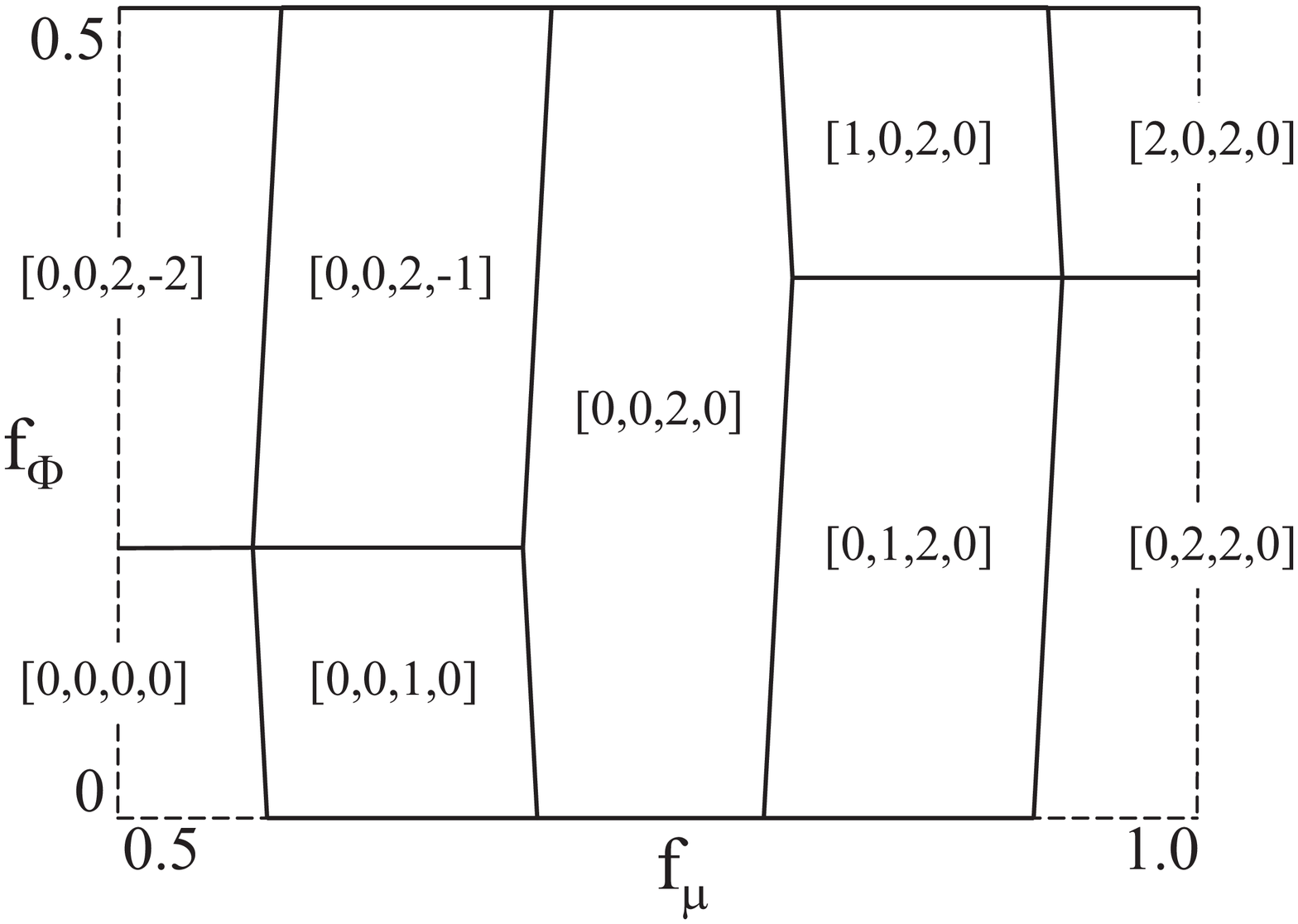}}
    \put(7.5,4.0){{\bf (b)}}
    \put(7.5,10.0){{\bf (a)}}
  \end{picture}
\vspace{-0.5cm}
\caption{The ground state of TNTs with $p=0$ (a) and $p=1$ (b) in terms of
the topological numbers $M_{d} = (M_{+}, M_{-})$ (a) and $M_{\alpha d} =
[M_{++}, M_{+-}, M_{-+}, M_{--}]$ (b). The ground state changes at the solid
lines. We choose $K_{++} = 0.2$. Inset: Electronic states near the crossing
point $\alpha = 1$ of the spectrum for TNTs with $p=0$ (circles) $-1$
(squares) $1$ (triangles) at $f_{\Phi} = 0$. The filling of the electronic
states with the energy $E \leq 2\pi v_{F}/3L$ at the branch $\alpha, s, d$
corresponds to $M_{\alpha s d} = 0$. The plot for $\alpha = -1$ can be
obtained by symmetry with respect to $k = 0$. }
\label{fig1}
\end{figure}

\begin{eqnarray}
A_{++} &=&1+\frac{4\bar{V}(0)}{\pi v_{F}}-\frac{\Delta V(0)}{4\pi v_{F}}-%
\frac{V_{+}(2K)}{2\pi v_{F}},  \label{App} \\
A_{\delta \nu } &=&1-\frac{\Delta V(0)}{4\pi v_{F}}-\delta \frac{V_{+}(2K)}{%
2\pi v_{F}},  \label{Adn} \\
B_{\delta \nu } &=&1+\frac{\Delta V(0)}{4\pi v_{F}}+\delta \frac{V_{-}(2K)}{%
2\pi v_{F}}.  \label{Bdn}
\end{eqnarray}
Here $V_{+}(q)$ and $V_{-}(q)$ are the amplitudes of intrasublattice and
intersublattice electron scattering with momentum transfer $q=0,2K$. The
forward scattering ($q=0$) has the strongest amplitude, $\bar{V}%
(0)=2e^{2}\ln (R_{s}/R)/\kappa $, where $\kappa $ is an effective dielectric
constant of the media (an estimate \cite{Egger} for the parameters of the
experiment \cite{Tans} gives $\kappa =1.4$), $R$ is the radius of the
nanotube, and $R_{s}\simeq \min (L/\pi ,D)$ characterizes the screening of
the Coulomb interaction due to a finite length $L$ of the TNT and/or the
presence of metallic electrodes at a distance $D$ \cite{Kane}. The
amplitudes $\Delta V(0)$ and ${V}_{+}(2K)$ decay as $1/R$ for $R\gg a$,
whereas $V_{-}(2K)\ll \min [\Delta V(0),{V}_{+}(2K)]$ due to the $C_{3}$
symmetry of the graphite lattice. Numerical evaluation for $R\gg a$ gives 
\cite{YoshOdi} $\Delta V(0)=0.21$, ${V}_{+}(2K)=0.60$, ${V}%
_{-}(2K)=9.4\times 10^{-4}$ in units of $ae^{2}/2\pi \kappa R$ (the
scattering amplitudes $\Delta V(0),{V}_{\pm }(2K)$ increase with decreasing
the localization radius $a_{0}$ of $p_{z}$ orbital; this phenomenological
parameter of the model \cite{Egger} is chosen as $a_{0}=a/2$).

Perturbation theory with respect to the backscattering and the
sublattice-dependent part of the forward scattering of electrons is
applicable if \cite{comm4} $\max [\Delta V(0),{V}_{+}(2K)]\ll 2\pi v_{F}$,
see Eqs. (\ref{Adn}), (\ref{Bdn}). This condition is equivalent to $N\equiv
2\pi R/\sqrt{3}a\gg c$, with $c\approx 0.1$ for the parameters listed above,
which is safely fulfilled for generic SWNTs with $N=10.$ The perturbation
splits degenerate electronic states $n,n^{\prime }$ belonging to the same
unperturbed energy level $i$. The splitting occurs already in first order
and can be estimated from the secular equation, $\det |V_{nn^{\prime
}}-E_{i}\delta _{nn^{\prime }}|=0$. The unperturbed states $n$ are
characterized by the topological numbers $N_{\delta \nu },\ J_{\delta \nu }$
and by the quantum state $|...\rangle _{b}$ of bosonic modes. Only the
vacuum state $|0\rangle _{b}$ has to be considered at low temperatures $T\ll
v_{F}/L$. The diagonal matrix elements $V_{nn}$ correspond to the energies
of the topological excitations (\ref{Htop}) (we will drop the constant
energy shift due to the renormalization of bosonic term (\ref{H})). The
topological and bosonic parts of non-diagonal matrix elements $V_{nn^{\prime
}}$ (\ref{Vf}), (\ref{Vb}) can be evaluated using the relations, $%
e^{il\theta _{\delta \nu }^{(0)}}|J_{\delta ^{\prime }\nu ^{\prime }}\rangle
=|J_{\delta \nu }+l\rangle \delta _{\delta \delta ^{\prime }}\delta _{\nu
\nu ^{\prime }}$, $e^{il\phi _{\delta \nu }^{(0)}}|N_{\delta ^{\prime }\nu
^{\prime }}\rangle =|N_{\delta \nu }+l\rangle \delta _{\delta \delta
^{\prime }}\delta _{\nu \nu ^{\prime }}$, and $\langle 0|e^{2i\tilde{\theta}%
_{\delta \nu }}|0\rangle _{b}=(2\pi \tilde{a}/L)^{K_{\delta \nu }}$, $%
\langle 0|e^{2i\tilde{\phi}_{\delta \nu }}|0\rangle _{b}=(2\pi \tilde{a}%
/L)^{1/K_{\delta \nu }}$. As a result we obtain

\begin{eqnarray}
&&\langle \vec{N},\vec{J}|\int_{-L/2}^{L/2}dx\cos 2\theta _{\delta \nu }\cos
2\theta _{\delta ^{\prime }\nu ^{\prime }}|\vec{N}^{\prime },\vec{J}^{\prime
}\rangle   \nonumber \\
&=&\left( \frac{L}{4}\right) \left( \frac{2\pi \tilde{a}}{L}\right)
^{K_{\delta \nu }+K_{\delta ^{\prime }\nu ^{\prime }}}  \label{matrelt} \\
&&\times \sum_{p,p^{\prime }=\pm }\delta _{J_{\delta \nu },J_{\delta \nu
}^{\prime }+2p}\delta _{J_{\delta ^{\prime }\nu ^{\prime }},J_{\delta
^{\prime }\nu ^{\prime }}^{\prime }+2p^{\prime }}\delta ...\delta
_{pN_{\delta \nu }^{\prime }+p^{\prime }N_{\delta ^{\prime }\nu ^{\prime
}}^{\prime },0},  \nonumber
\end{eqnarray}
and similar expressions for the other matrix elements. Here $\delta ...$
denotes that all topological numbers different from $J_{\delta \nu }$, $%
J_{\delta ^{\prime }\nu ^{\prime }}$ should be equal for the initial and
final states. The last term stems from the integration over $x$ in Eqs. (\ref
{Vf}), (\ref{Vb}). It produces an additional constraint on the topological
numbers, which can be traced back to the conservation of momenta of two
scattering electrons. The topological constraint has a somewhat different
form, $\delta _{pN_{_{--}}^{\prime }+p^{\prime }J_{_{--}}^{\prime
}+2pp^{\prime },0}$, for the last term in Eq. (\ref{Vb}), which contains two
field operators for the same $(--)$ sector. Let us note that the
non-diagonal part (\ref{Vf}), (\ref{Vb}) of the perturbation does not
contain matrix elements in the $(++)$ sector. For this reason, perturbed
ground states are characterized by a well defined topological number $J_{++}$
(and $N_{++}$) which determines the persistent current (\ref{Ip}) at zero
temperature.
Perturbed ground states are shown in Fig. 2. At not very small magnetic
flux, the perturbation lifts the spin degeneracy of two-electron (or
two-hole) ground states favoring spin aligned configurations (like $(2,0)$
in Fig. 2). With decreasing magnetic flux, the ''many-particle'' ground
states (with $2N_{++}=2...6$ $%
\mathop{\rm mod}%
$ $8$) experience reconstruction, so that
\begin{figure}[hbt]
\epsfxsize=\columnwidth\epsfbox{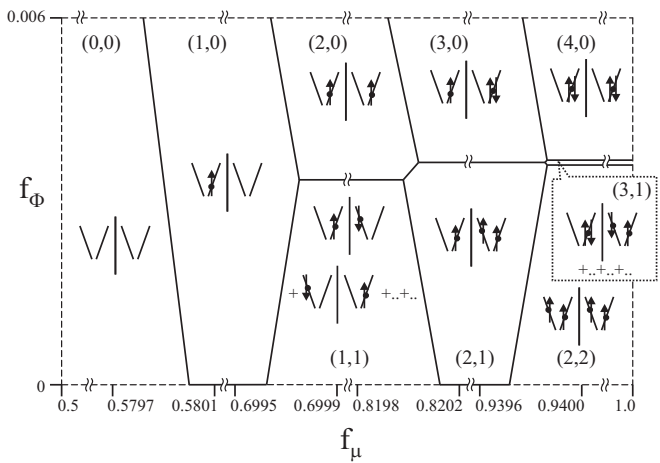}
\vspace{15pt}
\caption{The fine structure of the ground state for TNT with $p=0$. The
numbers $M_{+}, M_{-}$ of right- and left-movers are given in brackets. The
parameters $\Delta V(0),{V}_{\pm }(2K)$ are listed in the text below Eq. (\ref{Bdn}),
$K_{++} = 0.2$, and $\tilde{a} = 1/K$. Quantum states in a coherent
superposition are denoted by double dots. }
\label{fig2}
\end{figure}
\noindent
both the spin {\em and} orbital
configurations are changed. The reconstruction is observable as a jump of
the persistent current due to the change in numbers of right and left
movers. The increase of the kinetic energy of new ground states is
accompanied by the build-up of many-electron correlations, which minimizes
the total energy.

\smallskip For the states $(2,0)$, $(2,1)$, and $(2,2)$ in Fig. 2 the
electron spins are parallel, which is a signature of the exchange
interaction. The non-diagonal terms (\ref{Vf}), (\ref{Vb}) of the
Hamiltonian do not mix degenerate electron configurations corresponding to
each of these states. Let us note that the spin aligned ground states have
been presumably detected in very recent experiments \cite{Tans2} on
individual linear SWNTs, albeit the data differs substantially from the
results \cite{Cobden} on ropes of SWNTs.

The situation is different for the ground states $(1,1)$ and $(3,1)$ (Fig.
2). Each state represents a coherent superposition of $4$ configurations
with antiparallel spins, which has the lowest energy due to the interbranch
electronic exchange allowed by the non-diagonal matrix elements (\ref{Vf}), (%
\ref{Vb}) of the Hamiltonian. The new ground states $(1,1)$, $(2,2)$ with
even number of electrons are stable with respect to a change of sign of the
magnetic flux. For this reason, TNTs are diamagnetic for even $N_{tot}$ and
paramagnetic for odd $N_{tot}$, in contrast to the result of the constant
interaction model.

In conclusion, we have generalized the bosonization formalism for the case
of TNTs and evaluated the persistent current in this system away from
half-filling. The pattern of the persistent current depends on the number of
elementary cells along the nanotube modulo 3. The overall pattern (Fig. 1)
corresponds to the constant interaction model, whereas the fine structure
(Fig. 2) can be explained in terms of electronic exchange correlations. Even
though a system with a fixed electro-chemical potential was considered, the
results for fixed number of particles can be easily obtained from Eq. (\ref
{Ip}) and Figs. 1, 2 by an appropriate choice of the electro-chemical
potential. A submicroamp persistent current should be observable in a few
micrometer long TNTs. The Umklapp scattering of electrons on the atomic
lattice (at half-filling), impurities, structural imperfections, twiston
phonons, etc. may suppress the persistent current and deserves further
analysis.

\strut

We would like to thank G.E.W. Bauer, C. Dekker, Yu.V. Nazarov, and U. Weiss
for useful discussions. The financial support of FOM is gratefully
acknowledged. This work is also a part of INTAS-RFBR 95-1305. One of us
(A.O.) acknowledges the kind hospitality at the University of Stuttgart.


\begin{references}
\bibitem{comm2}  Drawbacks of the analysis \cite{Egger} are discussed
in Ref. \cite{YoshOdi}
\bibitem{comm3}  This follows from our definition of $M_{\alpha sd}$ (see
inset of Fig. 1a): filling of eight single-particle states at (near) the
crossing points of the spectrum corresponds to $M_{\alpha sd}=0$, whereas $%
N_{tot}=4$.
\bibitem{comm4}  More precise condition, $\Delta \ll \max (v_{F}/L,T)$, can
be obtained from the analysis of the strong coupling point where excitations
are gapful. Using the estimate for the (maximum) gap\cite{Egger}, $\Delta
=Kv_{F}\exp (-p),$ $p=\pi v_{F}/\sqrt{2}V_{+}(2K)$ we obtain $\Delta \simeq
10^{-13}$ eV for (10,10) SWNT.
\end{references}
\end{document}